\documentclass[compsoc, conference, a4paper, 10pt, times]{IEEEtran}

\usepackage{cite}
\usepackage{amsmath,amssymb,amsfonts}
\usepackage{algorithmic}
\usepackage{graphicx}
\usepackage{textcomp}
\usepackage{xcolor}
\usepackage{balance}
\usepackage{booktabs}
\usepackage{adjustbox}
\usepackage[hidelinks]{hyperref}

\begin{document}

\title{A first look into Utiq: Next-generation cookies at the ISP level}


\author{\IEEEauthorblockN{1\textsuperscript{st} Ismael Castell-Uroz}
\IEEEauthorblockA{\textit{Universitat Politècnica de Catalunya} \\
Barcelona, Spain \\
ismael.castell@upc.edu}
\and
\IEEEauthorblockN{2\textsuperscript{nd} Pere Barlet-Ros}
\IEEEauthorblockA{\textit{Universitat Politècnica de Catalunya} \\
Barcelona, Spain \\
pere.barlet@upc.edu}}

\maketitle

\begin{abstract}
Online privacy has become increasingly important in recent years. While third-party cookies have been widely used for years, they have also been criticized for their potential impact on user privacy. They can be used by advertisers to track users across multiple sites, allowing them to build detailed profiles of their behavior and interests. However, nowadays, many browsers allow users to block third-party cookies, which limits their usefulness for advertisers. In this paper, we take a first look at Utiq, a new way of user tracking performed directly by the ISP, to substitute the third-party cookies used until now. We study the main properties of this new identification methodology and their adoption on the 10K most popular websites. Our results show that, although still marginal due to the restrictions imposed by the system, between 0.7\% and 1.2\% of websites already include Utiq as one of their user identification methods.
\end{abstract}



\section{Introduction}
Online privacy is an increasingly important concern in today's digital age, as more and more of our personal information is stored and transmitted through the internet. 
One of the main concerns is the amount of data that companies collect about users. From search engines to social media platforms, companies gather vast amounts of information about browsing habits, purchases, and interests.
Until now, third-party cookies have been the most commonly used solution to perform this kind of identification. These cookies allow websites to track the user's activity across multiple sites, collecting information about their interests and behavior. This information is often used to personalize advertising content and improve marketing strategies, which is why many companies rely on it for their business operations. However, as concerns over online privacy have grown, many browsers now block third-party cookies by default (e.g., ~\cite{firefox_builtin, safari_builtin}), making them less useful for tracking user activity across the web.

In this context, Deutsche Telekom, Movistar, Vodafone, and Orange, four of the biggest telecommunications companies in Europe, have recently developed \textit{Utiq}, their own identification system with a special focus on privacy and transparency, to replace third-party cookies.
Utiq works by identifying the Internet connection tied to the user and generating from it a pair of anonymized tokens that can be used by companies for marketing and advertising purposes.
In this paper, we explore the main properties of Utiq and evaluate the level of adoption of this new system after six months of its inception.
To the best of our knowledge, no previous work has studied this new identification methodology.
Moreover, we also study if websites using Utiq also use other web tracking methods that, combined with Utiq's identification tokens, may be used to improve their user profiling systems, which would render the new system as controversial as third-party cookies.
Our results show that, although still marginal, six months after the creation of the service and depending on the country, between 0.7\% and 1.2\% of the 10k most popular websites include Utiq as one of their identifying processes. However, we also demonstrate that all of them complement this identification system with more intrusive methods, such as canvas or font fingerprinting~\cite{bujlow_survey_2017}, not resulting in a real privacy improvement over the third-party cookies that Utiq substitutes.

The rest of the paper is organized as follows: Section \ref{related_work} reviews the most relevant related work. Section \ref{utiq} describes the fundamentals of Utiq. Section \ref{study} presents the study of its level of adoption, and Section \ref{discussion} discusses some considerations about tracking as well as some of the challenges that Utiq must face in order to get a wider adoption. Lastly, Section \ref{conclusions} concludes the paper.

\section{Related work}
\label{related_work}
Web tracking has become an increasingly significant issue in recent years due to its potential impact on user privacy. One of the most prevalent web tracking methods has been third-party cookies. These files can be used to identify users across multiple websites. While they have been widely used for years, they have also been criticized for their potential impact on user privacy as they allow companies to build detailed profiles of their users behavior and interests. As such, there are many works in the literature studying the cookie ecosystem and its implications. In~\cite{leon_why_2012}, Leon et al. were one of the first to study the impact on usability of different tools that block cookies to improve the privacy of the user. Englehard et al. study in~\cite{englehardt_cookies_2015} how third-party cookies can be used to link visits from the same user to different websites even if the user's IP address is different. In~\cite{papadopoulos_cookie_2019}, Papadopoulos et al. study the proliferance of cookie syncing, a technique that allows different and not related third-party companies to sync the information about their users cookies in order to identify them even in services that are not self-owned.

In response to concerns about web tracking and user privacy, many websites have implemented privacy policies that outline how they collect and use data about their users. 
While privacy policies can help users understand what data is being collected about them, they are not always sufficient to ensure user privacy. 
To address these issues, some countries have implemented legal frameworks aimed at protecting user privacy online. For example, the European Union's General Data Protection Regulation (GDPR~\cite{gdpr}) requires websites to obtain explicit consent from users before collecting certain types of data. However, these laws are not always effective at preventing all forms of web tracking. Thus, many browsers have followed a more simple but effective way of limiting this tracking behavior: blocking all third-party cookies by default on all websites (e.g., Firefox~\cite{firefox_builtin}, Safari~\cite{safari_builtin}). Even Google Chrome, the most widely used browser, is currently progressively introducing this measure~\cite{google_builtin}.

However, most online companies rely on this kind of information for their business models. Thus, many of them are choosing to use more intruding techniques, such as browser or device fingerprinting, to obtain personal data. These techniques identify the device being used to browse the web by obtaining as much information as possible about it in order to create identifying fingerprints. We refer the interested reader to~\cite{bujlow_survey_2017} for a comprehensive survey with information about this kind of web tracking mechanism. On the other hand, works like~\cite{nikiforakis_cookieless_2013} from Nikiforakis et al. or~\cite{englehardt_online_2016} from Englehardt et al. have found a rise in the use of this kind of technique during the last years. Castell-Uroz et al. in~\cite{castell-uroz_tracksign_2021} quantify in more than 95\% of the total the number of websites that use fingerprinting or similar transparent tracking methods online. Although the research community has presented many works to automatically find and block this kind of web tracking by multiple means (e.g.,~\cite{iqbal_adgraph_2020, wu_machine_2016, acar_fpdetective_2013, kalavri_like_2016, le_towards_2017, ikram_towards_2016, li_trackadvisor_2015, yu_tracking_2016, metwalley_unsupervised_2015, gugelmann_automated_2015, castell-uroz_astrack_2023}), the difficulty of adoption of those techniques makes this still an open problem.

In this paper, we examine Utiq, a different approach taken by the some of the most important Internet service providers in Europe to substitute the use of third-party cookies. Utiq is based on the identification of the user by means of a set of anonymized tokens generated directly from the information about the connection of the user.

\begin{table}
\centering
\caption{Utiq compliant ISPs}
\label{tab:isps}
  \begin{adjustbox}{max width=0.44\textwidth}
  \small
  \begin{tabular}{p{0.3\textwidth}p{0.2\textwidth}}
  \hline
  \textbf{Country} & \textbf{ISP}                                                                   \\ \hline
  France  & \begin{tabular}[c]{@{}l@{}}Orange\\ Bouygues Telecom\\ SFR\end{tabular}                 \\ \hline
  Germany & \begin{tabular}[c]{@{}l@{}}Deutsche Telekom\\ Vodafone\\ Congstar\\ Fraenk\end{tabular} \\ \hline
  Spain   & \begin{tabular}[c]{@{}l@{}}Movistar\\ Orange\\ Jazztel\\ Simyo\end{tabular}             \\ \hline
  \end{tabular}
  \end{adjustbox}
\end{table}

\section{Utiq}
\label{utiq}
Until recently, online companies relied heavily on third-party cookies for marketing and advertisement purposes. However, many browsers now block these cookies by default (e.g., Firefox~\cite{firefox_builtin}, Safari~\cite{safari_builtin}), with Google Chrome expected to do so in 2024~\cite{google_builtin}. In response, Utiq~\cite{utiq}, a joint venture between Deutsche Telekom, Movistar, Vodafone, and Orange, four of the major European telecommunication operators, has developed its own digital user identification platform as an alternative to third-party cookies. The platform reportedly focuses on privacy and transparency, always asking for users consent before their identification.
While currently it is only available in Germany, Spain, and France, the company plans to expand its services to the United Kingdom and Italy in 2024. Table 1 provides a comprehensive list of internet service providers that allow user identification through Utiq's technology, including the main companies and their affiliates in these three countries.

\begin{figure}
  \centering
  \includegraphics[width=0.44\textwidth]{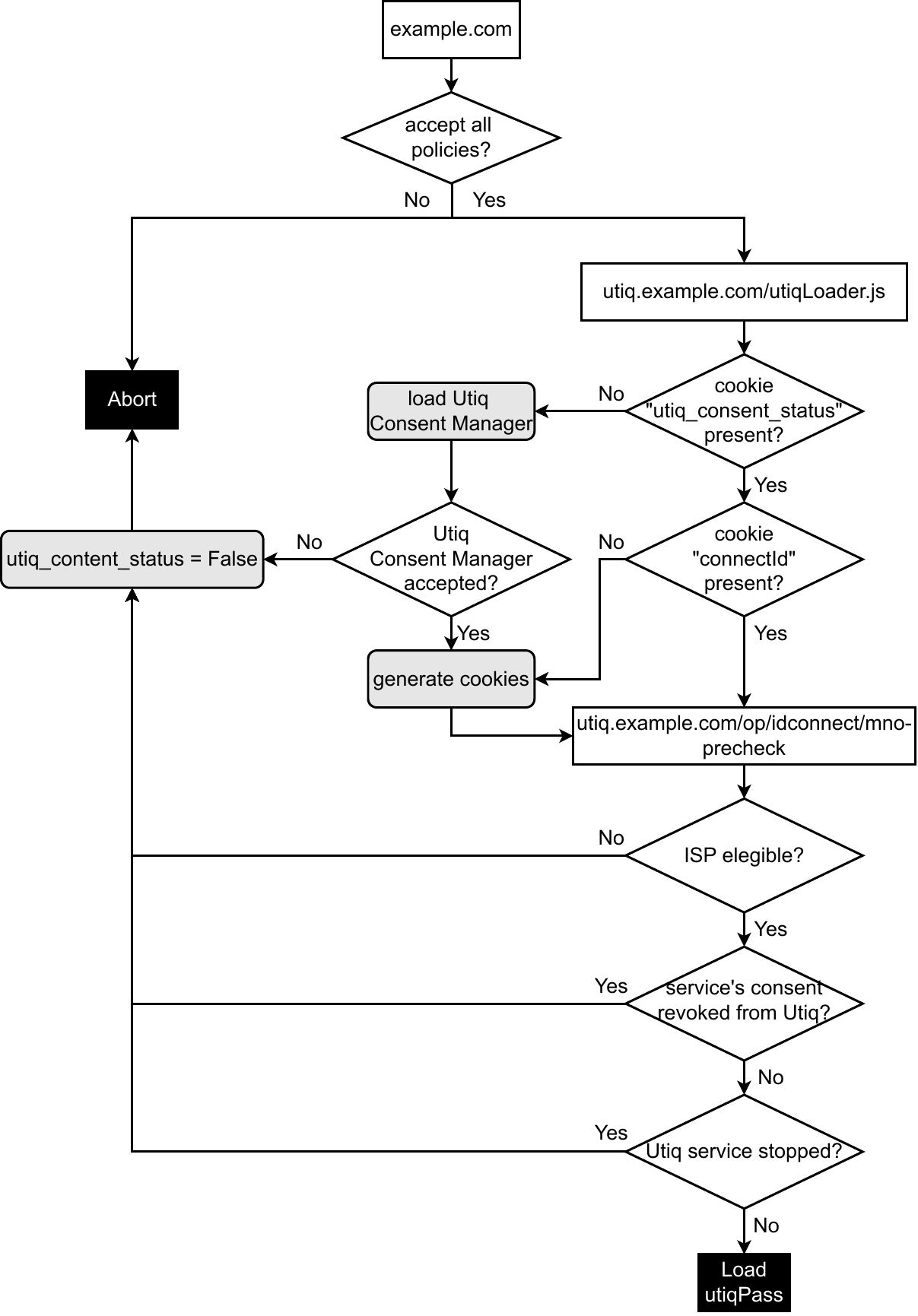}
  \caption{Utiq diagram}
  \vspace{-0.4cm}
  \label{fig:diagram}
\end{figure}

Utiq functions by acquiring what they refer to as a network signal called \textit{consentpass}. The consentpass is an identifier based on the user's Internet connection details and tied to the real user behind it. It is then used to generate two anonymized and distinct signals, the \textit{martechpass} and \textit{adtechpass}. These signals serve different purposes within the advertising ecosystem. The martechpass focuses on marketing objectives and is generated individually for each brand or publisher that incorporates Utiq services. The marketing ecosystem seeks to identify users in order to deliver personalized messaging to engage with customers (e.g., social media, email, or interactive content marketing). From Utiq's perspective, a single publisher can share the martechpass between multiple services as long as the user explicitly consents to sharing this kind of information. In contrast, the adtechpass is employed for programmatic advertising within the adtech ecosystem. The advertising ecosystem aims to find the most effective way to promote a product or service through paid or "sponsored" channels. The adtechpass is used to uniquely identify users and helps to prevent inefficiencies such as displaying the same advertisement repeatedly. This allows advertisers to optimize their expenditures.

Utiq offers a single location called \textit{Consenthub} for managing all online service permissions. Within this platform, users can fully deactivate Utiq service usage for up to a year. Although websites may continue requesting user consent to utilize Utiq services during this timeframe, the identification process will not proceed. Afterward, the status resets, and users can either approve specific service consents or temporarily disable Utiq again for another year. As Utiq is a system aimed at being deployed in European countries, Consenthub complies with all the regulations introduced in the GDPR~\cite{gdpr}.

Fig. \ref{fig:diagram} showcases Utiq's identification methodology. Initially, to retrieve the network signal and generate the two identifying tokens, users must first \textbf{accept all} privacy policies on the main website. The website then downloads a JavaScript file (utiqLoader.js) responsible for checking that all the required privacy consent conditions are met by the user. This file is provided through the "utiq" subdomain of the primary website (e.g., \textit{utiq.example.com/utiqLoader.js}). This subdomain must be redirected via a CNAME record to the URL \textit{frontend.prod.utiq-aws.net}, where the latest version of the file can be accessed. Once downloaded and executed, the script checks for two first-party cookies called ``utiq\_consent\_status'' and ``connectId.'' If the first cookie is not available, a new consent manager is loaded to obtain user consent for generating and using Utiq identifiers. The second cookie only exists if the user has already given consent to load it and identifies the network signal used to generate the consentpass. This cookie mainly serves to decrease the burden on ISPs of checking connection details every time. Both cookies are valid for 90 days, after which the user must explicitly give consent again through Utiq's consent manager. Once the connection details are known, the JavaScript checks the connection by means of a call to ``utiq.example.com/op/idconnect/mno-precheck'' and saves it inside a new object in the browser's local storage called \textit{utiqEligibility}. This object identifies if the ISP being used to open the website is eligible for using the service or not, or, in other words, if it is one of the ISPs present in table~\ref{tab:isps}. It also verifies if the specific website being opened has already been revoked from Utiq's Consenthub or if the Utiq service has been stopped altogether. In both cases, it will halt the loading process of the two identification tokens. Only when the user has given explicit consent to both the main online service and Utiq's service and the connection is done through one of the compliant ISPs will both tokens be available through a new object in the browser's local storage called \textit{utiqPass}.

Finally, Utiq also provides a way to programmatically integrate Utiq's adtechpass to an SSP (Supply-Side Platform) by means of the \textit{prebid.js} library. SSPs are interfaces that publishers use to automatically sell their remnant (ad space not sold through direct deals) inventory. The prebid.js library includes not only Utiq but also many other user identification services such as Merkle, Criteo, or Quantcast. Regarding the programmatic connection to the DSP (Demand-Side Platform), Utiq has an exclusivity agreement with AdForm~\cite{adform}, one of the most important DSPs. Thus, websites that use Utiq to automatically sell their advertisement space will only receive advertisements included in the portfolio of AdForm.

\section{Adoption study}
\label{study}
Utiq started operating in August 2023. With most browsers now blocking third-party cookies by default and Google's Chrome progressively introducing the same measure during 2024, this period should be decisive to determine the acceptance that this new system will have during the next few years. In this section, we study the current adoption of the system, almost six months after the starting date.

\subsection{Methodology}
Utiq is currently available only in three countries: France, Germany, and Spain. However, we studied the top 100k most popular websites from the Tranco List\footnote{Available at https://tranco-list.eu/list/24W99}~\cite{pochat_tranco_2018} generated on January 15th 2024, regardless the country where it is hosted, to discover possible foreign domains identifying European users. 
For each website in our dataset, we tried to directly load the URL "https://utiq.domain.com/utiqLoader.js", one of the requirements websites have to follow in order to use Utiq for their identification systems. The inspection was directly done by means of a combination of Python and the CURL tool. Moreover, the URL "https://utiq.domain.com/op/idconnect/mno-precheck" was also inspected. This address should return a \{“status”:“ok”\} message if the ISP supports Utiq or \{“status”:“Not found”\} otherwise. Only in the event that both URLs are accessible will we account for the domain as compliant with Utiq, as per their requirements.


\begin{table}
\centering
\caption{Utiq presence}
\label{tab:presence}
  \resizebox{0.44\textwidth}{!}{
  \small
\begin{tabular}{lccc}
\hline
\textbf{Country} & \textbf{Websites} & \textbf{Utiq compliant} & \textbf{Percentage} \\ \hline
\textbf{Total}   & 9316              & 83                      & 0.89\%              \\
\textbf{France}  & 2048              & 14                      & 0.68\%              \\
\textbf{Germany} & 3591              & 27                      & 0.75\%              \\
\textbf{Spain}   & 3641              & 42                      & 1.15\%              \\ \hline
\end{tabular}}
\end{table}

\subsection{Results}
Fortunately, all the domains using Utiq pertained to one of the three countries where Utiq is available. Thus, other countries, especially non-GDPR compliant ones, are not using the system yet. To inspect the adoption rate we selected the subset of domains pertaining to one of those three countries (about 10k websites) and checked Utiq prevalence for each of them.
Table~\ref{tab:presence} presents the overall obtained results for the subset of all those domains as well as those divided by country. Although still quite marginal, there are differences in adoption between the three countries, with Spain leading with a rough 1.2\%, while Germany and especially France show less introduction with 0.75\% and 0.68\%, respectively. Interestingly, all the websites included in the list except three of them are online newspapers. From the other three websites, two are websites about cooking recipes, and the last one is ``atf-tagmanager.de'', a web tracking service that seems to have included Utiq's identifier in their identification processes. One website, ``actu.fr'', loaded the \textit{utiqLoader.js} file before getting the consent of the user in some of our visits. This goes against Utiq requirements and is possibly a configuration error. Finally, about 80\% of the websites including Utiq (66 out of 83) integrate programmatic selling processes via the \textit{prebid.js} library. Appendix~\ref{appendix} contains the complete list of websites compatible with Utiq services, as well as their country and their rank within the Tranco list.

Although Utiq emphasizes their private and transparent identification methods, the process is similar to that of actual third-party cookies, with a unique token generated per user available for the internal website identification systems. In fact, comparing both approaches, Utiq may be considered more intrusive as the tokens are based on completely unique parameters that can precisely identify the user through time, which is impossible with regular cookies that the user can periodically clean. Moreover, websites including Utiq services can also use other web tracking systems to complement their identification processes to better profile the user, and here Utiq can represent an important concern by following users over time.

\begin{figure}
  \centering
  \includegraphics[width=0.46\textwidth]{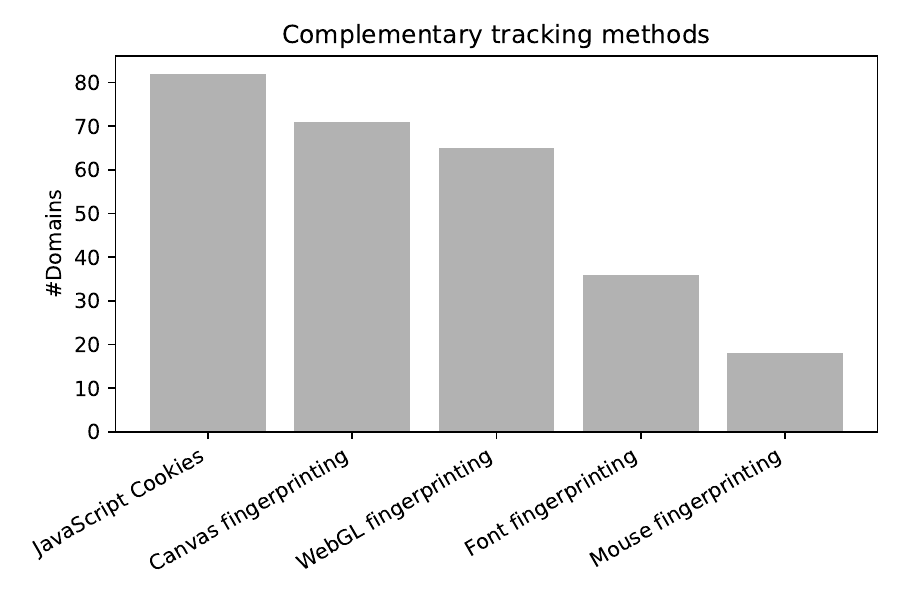}
  \caption{Complementary tracking methods}
  \vspace{-0.4cm}
  \label{fig:tracking}
\end{figure}

To study this situation, we explored other tracking methods that Utiq-compliant websites may also be using. This information was collected using ORM~\cite{castell-uroz_tracksign_2021}, a tool to perform studies about web tracking on a given population of websites. We inspected not only the homepage but also all the links contained in it that pertain to the same domain. Figure~\ref{fig:tracking} presents the distribution of other tracking methods currently included in the 83 websites detected using Utiq. As shown, all the websites include other tracking techniques running in the background that can complement the information obtained for each individual. Moreover, some of those systems are very intrusive tracking methods, such as canvas or font fingerprinting. Thus, we can conclude that currently all the websites complement the user information with other web tracking sources to obtain a more precise profile of the user.

\section{Discussion}
\label{discussion}
Many actors have been exploring ways to enter the online advertising ecosystem for years. The reason for this is simple: online advertisements are a major source of income for many online companies. ISPs run the core network where the Internet runs but, until now, they did not try to obtain economic profit from the advertisement ecosystem running inside. Utiq is a step into this direction, taking profit from the end of third-party cookies. It allows them to capitalize on the growth in online advertising by leveraging their unique position as the gateway to the Internet through the offer of precise user identification, which is highly valuable to businesses.
Nevertheless, they offer the service, emphasizing its focus on privacy and transparency by explicitly asking for the user's consent in all cases and not saving personal information about the browsing history. In practice, privacy-wise, the combination of their identification method with other web tracking systems can not only be as controversial as the third-party cookies, but even more because of the possibility of precisely identifying the user over time.

On paper, Utiq seems like a good approximation to substitute the third-party cookies used by many websites to identify the user. However, the adoption of the system after six months is still very marginal, containing only a few dozen websites. Thus, there must be a reason that holds other websites back from integrating Utiq identification into their services. Although a speculation, the fact that the user must accept two independent consent managers, one from the main website and the other from Utiq's service itself, may interfere with the adoption of the system, as many websites were also reluctant to introduce the cookie consent manager forced by the GDPR back in 2018 to not lose users. On the other hand, the exclusive agreement with AdForm~\cite{adform} may also be impacting their introduction, as there are many other advertisers in the market that cannot compete for their bidding space because of not being included.

\section{Conclusions and Future Work}
\label{conclusions}
In this work, we took a first look at Utiq, a new identification method proposed by some of the major internet service providers in Europe. The identification is done based on two anonymized tokens extracted from a unique identifier of the user, the internet connection. Our results show still a marginal adoption of less than 1\% after six months of operation. Although they highly emphasize the private and transparent nature of the system, using this methodology, companies can unambiguously identify the user over time. By itself, the system could be a privacy-friendly approach for publishers and advertisers, but when used in combination with other web tracking methods, it may allow companies to create even more precise user profiling. Our study shows that currently all websites including Utiq as one of their identification methods also use other complementary web tracking systems, with more than 80\% of them using intrusive mechanisms such as fingerprinting algorithms. Our future work includes studying if other identification propagation techniques, such as \textit{cookie syncing}, may be used with this kind of unambiguous identification.

\section*{Acknowledgements}
This work was supported by the CHISTERA grant CHIST-ERA-22-SPiDDS-02 corresponding to the GRAPHS4SEC project (reference nº PCI2023-145974-2) funded by the Agencia Estatal de Investigación through the PCI 2023 call.

\bibliographystyle{ieeetr}
\balance
\bibliography{references}

\newpage
\appendices
\section{Utiq compliant websites}
\label{appendix}

\begin{minipage}{0.92\textwidth}\centering
  \resizebox{\textwidth}{!}{
  \footnotesize
  \begin{tabular}{p{0.25\textwidth}p{0.5\textwidth}p{0.25\textwidth}}
\hline
\textbf{Tranco rank} & \textbf{Domain}             & \textbf{Country} \\ \hline
786                  & lefigaro.fr                 & France           \\
841                  & bild.de                     & Germany          \\
936                  & elmundo.es                  & Spain            \\
1018                 & marca.com                   & Spain            \\
1065                 & ouest-france.fr             & France           \\
1259                 & merkur.de                   & Germany          \\
1339                 & actu.fr                     & France           \\
1506                 & marmiton.org                & France           \\
2385                 & welt.de                     & Germany          \\
2888                 & faz.net                     & Germany          \\
2956                 & focus.de                    & Germany          \\
3328                 & abc.es                      & Spain            \\
4454                 & eldiario.es                 & Spain            \\
4772                 & elperiodico.com             & Spain            \\
5316                 & sudouest.fr                 & France           \\
5547                 & fr.de                       & Germany          \\
5693                 & sport.es                    & Spain            \\
6396                 & elcorreo.com                & Spain            \\
6976                 & tz.de                       & Germany          \\
7059                 & hna.de                      & Germany          \\
7836                 & voici.fr                    & France           \\
7850                 & lachainemeteo.com           & France           \\
9058                 & programme.tv                & France           \\
10490                & expansion.com               & Spain            \\
11062                & wiwo.de                     & Germany          \\
12455                & utopia.de                   & Germany          \\
13732                & farodevigo.es               & Spain            \\
14685                & mopo.de                     & Germany          \\
14992                & lasprovincias.es            & Spain            \\
15197                & lne.es                      & Spain            \\
15302                & diariovasco.com             & Spain            \\
15409                & levante-emv.com             & Spain            \\
16089                & elcomercio.es               & Spain            \\
25772                & aufeminin.com               & France           \\
29822                & ideal.es                    & Spain            \\
30343                & laverdad.es                 & Spain            \\
30384                & diariosur.es                & Spain            \\
30903                & eldiariomontanes.es         & Spain            \\
33551                & wa.de                       & Germany          \\
34097                & caminteresse.fr             & France           \\
34169                & fnp.de                      & Germany          \\
35965                & epe.es                      & Spain            \\
36443                & diariodemallorca.es         & Spain            \\
36637                & kreiszeitung.de             & Germany          \\
36949                & hoy.es                      & Spain            \\
37321                & elnortedecastilla.es        & Spain            \\
38758                & informacion.es              & Spain            \\
39119                & autoplus.fr                 & France           \\
39152                & elperiodicodearagon.com     & Spain            \\
40657                & promiflash.de               & Germany          \\
41891                & laopiniondemurcia.es        & Spain            \\
42416                & eldia.es                    & Spain            \\
42919                & laprovincia.es              & Spain            \\
43756                & laopiniondemalaga.es        & Spain            \\
45457                & larioja.com                 & Spain            \\
46415                & diariocordoba.com           & Spain            \\
47312                & telva.com                   & Spain            \\
47428                & elcorreogallego.es          & Spain            \\
47496                & elperiodicoextremadura.com  & Spain            \\
47714                & ingame.de                   & Germany          \\
47807                & op-online.de                & Germany          \\
49123                & charentelibre.fr            & France           \\
49246                & laopinioncoruna.es          & Spain            \\
50017                & diariodeibiza.es            & Spain            \\
50195                & elperiodicomediterraneo.com & Spain            \\
51510                & laopiniondezamora.es        & Spain            \\
53081                & superdeporte.es             & Spain            \\
54247                & maison-travaux.fr           & France           \\
55751                & diaridegirona.cat           & Spain            \\
55823                & larepubliquedespyrenees.fr  & France           \\
56452                & leonoticias.com             & Spain            \\
60603                & come-on.de                  & Germany          \\
64445                & buzzfeed.de                 & Germany          \\
65374                & regio7.cat                  & Spain            \\
65584                & soester-anzeiger.de         & Germany          \\
71058                & eatbetter.de                & Germany          \\
71971                & bw24.de                     & Germany          \\
73728                & landtiere.de                & Germany          \\
73831                & 24vita.de                   & Germany          \\
82417                & einfach-tasty.de            & Germany          \\
90871                & lavozdigital.es             & Spain            \\
95878                & atf-tagmanager.de           & Germany          \\
97222                & 24auto.de                   & Germany          \\ \hline
\end{tabular}}
\end{minipage}

\end{document}